\documentclass{psa}

\usepackage{amsmath}
\usepackage{subcaption}

\begin{document}

\doival{10.1017/psa.2026.10255}
\jnlPage{}{}
\jnlDoiYr{}
\lefttitle{Are Widely Known Findings Easier to Retract?}

\papertitle{RESEARCH ARTICLE}

\title{Are Widely Known Findings Easier to Retract?}

\author{Shahan Ali Memon$^{1}$, Jevin D. West$^1$ and Cailin O'Connor$^2$}

\affil{$^{1}$Information School, University of Washington, $^{2}$Department of Logic and Philosophy of Science, UC Irvine \\
      $^\ast$Corresponding author: Cailin O'Connor;
      \email{cailino@uci.edu}}

\received{XX XX XXXX}
\revised{XX XX XXXX}
\accepted{XX XX XXXX}

\begin{abstract}
Failures of retraction are common in science.  Why do they occur?  And what determines whether a retraction is successful?  We use data from citation records and Altmetrics to test proposed answers to these questions.  \citet{lacroix2021dynamics} employ network models to argue the social spread of information helps explain failures of retraction.  One prediction is that widely known results, surprisingly, should be easier to retract, since their retraction is more relevant.  Our results support this conclusion.  We find highly cited papers show more significant reductions in citation after retraction and garner more attention to their retractions as they occur.  
\end{abstract}

\maketitle

\section{Introduction}

Scientific retraction is not always effective.  Across many fields retracted papers are cited approvingly after retraction, sometimes at surprisingly high rates.\footnote{There are now hundreds of studies yielding similar findings across diverse academic disciplines and types of results.  For some examples see \citet{Bornemann-Cimenti-2016, neale2010analysis, pfeifer1990continued, van2016propagation, madlock2015lack, teixeira2017highly, szilagyi2022citation, shuai2017multidimensional, kuhberger2022self}.}  This poses a problem for scientific progress.  New studies draw on existing findings but cannot do so effectively if those findings are not reliable.  In the wake of the metascience movement, and under the influence of the Retraction Watch blog, there is increasing attention to this issue.

One relevant question is: Which findings are easier or harder to retract, and why? In previous work using models of social networks \citet{lacroix2021dynamics} predict that finding with wide reach should be easier to retract.  Humans tend to share information that is relevant.  While new scientific findings are broadly relevant, information about retraction is most relevant to those already misinformed about a finding. For this reason, information about the retraction of widely known findings spreads effectively in their models.

We test this modeling prediction using citation data and Altmetric data (tracking media attention).  In the first case, we predict a more significant drop in citations for highly cited papers after retraction.  We find support for this prediction.  More highly cited papers see significantly greater drops in citation count upon retraction.  They also see a larger relative drop in citation rate.  In the second case, we predict, and find, more online attention to highly cited papers during their retractions.

Our findings provide new insight into the process of retraction, and to the role of social aspects of knowledge in retraction.  They also demonstrate the usefulness of combining methods of inquiry---in this case simple social models and data analysis---to the study of philosophy of science and metascience.

In the next section, \ref{model}, we describe the model from \citet{lacroix2021dynamics}, motivating it with some relevant empirical findings.  We also describe our main predictions.  In section \ref{data} we describe the two data sets we make use of, as well as our methods of analysis.  In section \ref{results} we present our main results.  And section \ref{conclusion} provides relevant discussion.

\section{Model and Predictions}
\label{model}

There are myriad candidate explanations for failures of retraction---journals do not actively communicate about retraction, some scientific search engines fail to clearly identify retracted papers, and individual scientists are often hesitant to communicate about their retractions for reputational reasons.  Beyond these issues, features of human information sharing systems may prompt such failures as well.

\citet{lacroix2021dynamics} study the social dynamics of retraction on the empirically well-supported assumption that information spreads via social contact in human groups. They draw on SIR (susceptible, infected, recovered) models from epidemiology, which track the spread of infectious diseases, but are also used to study the spread of rumors or information in human groups \citep{Rogers-2012, Hayhoe-et-al-2017}.

Their basic model is a connected network of $N$ agents who can be in one of three belief states.  Neutral (susceptible) agents have never encountered a new, false scientific claim, false (infected) agents have encountered this claim and now hold a false belief, retracted (recovered) agents have encountered information about retraction and correctly believe the original claim to be false.  In each round of simulation, an agent is randomly paired with a network neighbor, and their belief states may influence each other.  Importantly information spread is asymmetric.  Neutral agents may learn false beliefs, and false agents may learn about a retraction.  Thus agent states only flow from neutral $\rightarrow$ false $\rightarrow$ retracted.  

A substantive assumption is that agents do not learn about retractions before learning false beliefs.  As noted, the authors justify this via appeal to the role of relevance in human communication---people tend to avoid sharing irrelevant information \citep{Grice-1975, sperber1986relevance, wilson2006relevance}.  Since retractions are typically only relevant to those who learned the original finding, information about them will not spread well absent this context.

A central result is that when agents do not share new information indefinitely, false beliefs persist in these models as an accident of history.  Some agents learn a false finding, and never happen to encounter its retraction. The authors explore conditions under which larger or smaller proportions of the population maintain permanent false beliefs.

\begin{figure}[ht]
\centering
  \includegraphics[width=.9\linewidth]{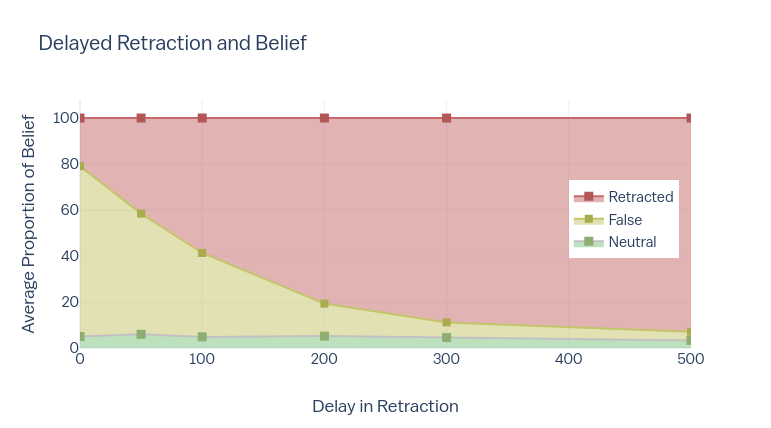}
\caption{\textbf{Delayed retractions are more effective}. Results for fully connected network of size $N=100$ where agents share newly learned beliefs for 200 time steps of simulation.}
\label{fig:model}
\end{figure}

Most important for us is the role of delay in retraction.  One might think that a retraction should be issued as soon as possible to prevent eventual false beliefs.  In their models, though, delayed retractions are issued in a network where many agents have already encountered the false belief.  The retraction can thus spread more readily, and is more successful. We present this effect in figure \ref{fig:model} which shows average final belief states in this model for different lengths of delay in retraction.\footnote{This figure uses original data from \citet{lacroix2021dynamics}.} 

The models just described are highly simplified and do not fully track real dynamics of information sharing.\footnote{For example, scientists do not always accept new results, and may spread objections or counter results.  Scientists also have graded levels of belief in some theories, and complex causal relations between beliefs \citep{sep-epistemology-social, golub2010naive, bala1998learning}.}  As such, they should not be taken to directly inform real scientific communities.  But we can use them to generate testable predictions about causal processes that might occur in the real world. 

In the models delay improves retraction because delay leads false findings to spread more widely. But in the real world the importance of the research, publication venue, status of the authors, their social sway, and chance events all determine the degree to which a finding is widely shared in a community. Thus, the relevant prediction is that widely known findings should be easier to retract, and that this retraction should generate more attention.  Specifically we predict that:

\begin{enumerate}
  \item Highly cited papers will see a more significant drop in citations after retraction (P1).
  \item Highly cited papers will receive relatively more Altmetric mentions during retraction (P2).
\end{enumerate}

In the case of the first prediction, we assume citation rate correlates with reach of a finding in a scientific community.  We also assume citations signal belief or trust in a piece of research.\footnote{In the discussion we address cases where these assumptions might fail.}  Of course, a popular finding may still be widely cited even after a relatively successful retraction compared to a finding that received no attention in the first place.  For this reason we look at two measures of drop in citation rate meant to track both the absolute drop and the drop as a percentage of expected future citations.\footnote{In this our predictions deviate from the model where widespread results will be less widely held after retraction full stop.  In the real world, unlike the model, many papers will never receive any attention.  It is unlikely that even successful retractions will lead widespread results to be cited by no one.}

As a secondary test, we also look at Altmetrics. During retraction, papers tend to see a bump in attention, presumably corresponding to the spread of information about the retraction~\citep{memon2025characterizing}.  We expect highly cited papers to see a more significant bump---corresponding to the model prediction that more relevant retractions will spread more readily in social networks.\footnote{Again of course in these models there are no Altmetrics explicitly represented. Instead they predict a high rate of social information spread, which Altmetrics should track.}

\section{Data, Methods, and Analysis}
\label{data}

To test prediction 1 (P1), we relied on two primary datasets: The first is Retraction Watch (RW), the most comprehensive publicly available repository of retracted research, comprising over 26,000 papers (as of the time of data acquisition) across approximately 5,800 journals and conferences \citep{watch2021tracking}. The second is the Microsoft Academic Graph (MAG), which provides metadata and citation networks for over 263 million scientific publications and captures the publishing trajectories of more than 271 million authors. For our analysis, we removed all records that were retracted in bulk (e.g., more than 1200 abstracts retracted by a single conference in 2011~\citep{mccook2018one}). We also only retain papers retracted between 1990 and 2015 to allow for 5 years of retraction window to study post-retraction outcomes. After filtering, we merged RW and MAG records based on DOI. For papers that did not find a DOI match, we used fuzzy matching\footnote{https://github.com/RobinL/fuzzymatcher} to match paper titles.\footnote{Fuzzy matching is a data matching method that connects data entities that are not exact matches if one compares character to character but are indeed the same entity (e.g.,``The cowbell" -``The cow-bell" ). There are various algorithms that are used to achieve fuzzy matching, such as Levenshtein distance, Jaccard index, etc.} This resulted in a total of 6,188 retracted papers, or the ``filtered dataset" henceforth.

\subsection{P1: Studying the change in citation post-retraction}
Our hypothesis posits that the effect of retraction on subsequent citation behavior is modulated by a paper's pre-retraction visibility, operationalized as the number of citations received prior to retraction. Specifically, we expect that papers with greater pre-retraction visibility experience a larger decline in citations post-retraction. 

To test this, we employ matching, which is a statistical method for assessing treatment effects (e.g., retraction). For each retracted paper in our filtered dataset, we identify one or more non-retracted papers matched exactly on: (a) publication year, (b) journal or conference, (c) scientific discipline, and (d) total number of citations received from publication up to the retraction year (inclusive). Scientific discipline labels are assigned based on the MAG field-of-study hierarchy using methodology from prior work by \citet{memon2025characterizing}. This procedure yields an analytical sample of 5,256 retracted papers, matched to 651,829 non-retracted control papers, with an average of 155, and a median of 27, matched controls per retracted paper. This matching allows us to make predictions for what the citations of retracted papers might have looked like had they not been retracted, and then compare the citations of matched retracted papers to these predictions.  Figure \ref{fig:percentdrop1} shows, conceptually, how this matching works.\footnote{There are limitations to this approach, which we address further in the discussion.} 

To quantify the impact of retraction across different levels of pre-retraction visibility, we define two outcome metrics, operationalizing the prediction in different ways. As will become clear, each metric has limitations.  Both together present a fuller picture of how pre-retraction attention influences success of retraction.

\subsubsection{Outcome 1: Absolute post-retraction citation difference}
This outcome captures the absolute difference in post-retraction citation volume between a retracted paper and its matched control:
\[
\text{Outcome}_1 = \text{Post}_r - \text{Post}_m
\]

We limit our post-retraction citations to 5 years, which is typical for bibliometric analysis, to allow for a consistent window for all the papers regardless of their publishing year. Because retracted and control papers are exactly matched on their pre-retraction citation counts, this outcome can be interpreted as a simplified version of the classic difference-in-differences estimator:
\[
(\text{Post}_r - \text{Pre}_r) - (\text{Post}_m - \text{Pre}_m)
\]

with the pre-retraction terms effectively controlled for via matching. A more negative value indicates a larger real-world citation penalty for the retracted paper, relative to its matched counterpart.  We predict a larger absolute drop in future citations for papers highly cited before retraction.

One limitation of this outcome is that since pre-retraction citation rates can be dramatically different across papers, even a minor percentage drop in citation rates for highly cited papers can be quite large in terms of absolute citations compared to a large percentage drop for a paper with few citations.  For that reason we calculate a second metric.

\subsubsection{Outcome 2: Post-retraction citation ratio}
This outcome compares the post-retraction citations of retracted papers to their matched controls:
\[
\text{Outcome}_2 = \frac{\text{Post}_r}{\text{Post}_m}
\]

We estimate this using the log-transformed ratio of post-retraction citations as follows:

\[
\text{Outcome}_2 = \log(\frac{\text{Post}_r + \epsilon}{\text{Post}_m+ \epsilon}) 
\]
where \(\epsilon\) is a small constant (i.e., \(10^{-5}\) in our case) added for numerical stability to avoid division by zero. This tracks the relative, or percentage, change in citations for retracted papers versus their matches. In other words, it computes the proportional change in citation compared to what we would predict. Because citation counts are often highly skewed, we use a logarithm scale to compress extreme ratios and make comparisons more interpretable.  We predict larger drops in this outcome for papers more highly cited pre-retraction.

While the first metric arguably exaggerates the success of retraction for widely known papers, the second arguably does the opposite.  For a paper with one expected post-retraction citation, losing one citation amounts to a 100\% reduction.  A paper with 100 expected citations can lose 99 of them, and still not achieve that level of reduction. For this reason we think both metrics are useful in more fully understanding post-retraction citation reductions.
\\

The above outcomes are computed at the level of individual retracted paper–control paper pairs. For papers with more than 1 match, we take their average outcome. To evaluate whether the effect of retraction varies systematically with pre-retraction visibility, we stratify papers into citation groups based on their total citations prior to retraction. More concretely, citation groups were defined based on quantiles of the pre-retraction citation distribution across all retracted papers in the filtered sample. Specifically, we used the 25th, 50th, 75th, and 90th percentiles to divide papers into following visibility-based citation tiers: $[0,1), [1,9), [9,31), [31,inf)$. 

For each pre‑retraction citation group we summarise outcome 1 with the
\textit{mean} across all retracted–control pairs.
Outcome 2, which is the logarithm of the post‑retraction citation ratio, is summarised with the \textit{median}. Because the denominator
(post‑retraction citations of the matched control) can be close to zero,
the log‑ratio has a long right tail. The median therefore provides a robust measure of the typical effect, while we report the mean and maximum in the online appendix Table 2 to show the full spread of the distribution.\footnote{Available at \url{https://osf.io/kd3bw/?view_only=3c0cfc7d50a549588305e6575317f096}} 

\subsection{P2: Studying the association between pre-retraction citations and retraction-related attention}

Our second hypothesis posits that highly cited papers receive more retraction-related attention. To operationalize attention, we further merge another dataset, Altmetric, with our filtered RW-MAG sample using DOI. Altmetric tracks over 191 million mentions of research outputs across social media, blogs, news media, and knowledge repositories, covering more than 35 million scholarly works. Altmetric provides aggregate scores for the entire time period of the paper. To isolate attention related to retraction, we only look at a 12 month window surrounding the retraction date. Details on how the score was computed and validated are based on prior work by \citet{memon2025characterizing}. Along with the Altmetric score, which is calculated using weights over all mentions\footnote{For example, a newspaper mention is more weighted than a Bluesky post because the former usually correlates with more attention.}, we estimate a model of total mentions as a second outcome for this hypothesis.

To study the relationship between pre-retraction citations and attention, we estimate a linear regression model, controlling for the year of publication, number of years between publication and retraction, journal ranking, the reason for retraction, the number of authors on the paper, and subject area (as assigned by RW). After filtering for complete cases across all control variables, we retain a final analytical sample of 3,271 observations.

\section{Results}
\label{results}

\begin{figure}[!htbp]
\centering
  \includegraphics[width=\linewidth]{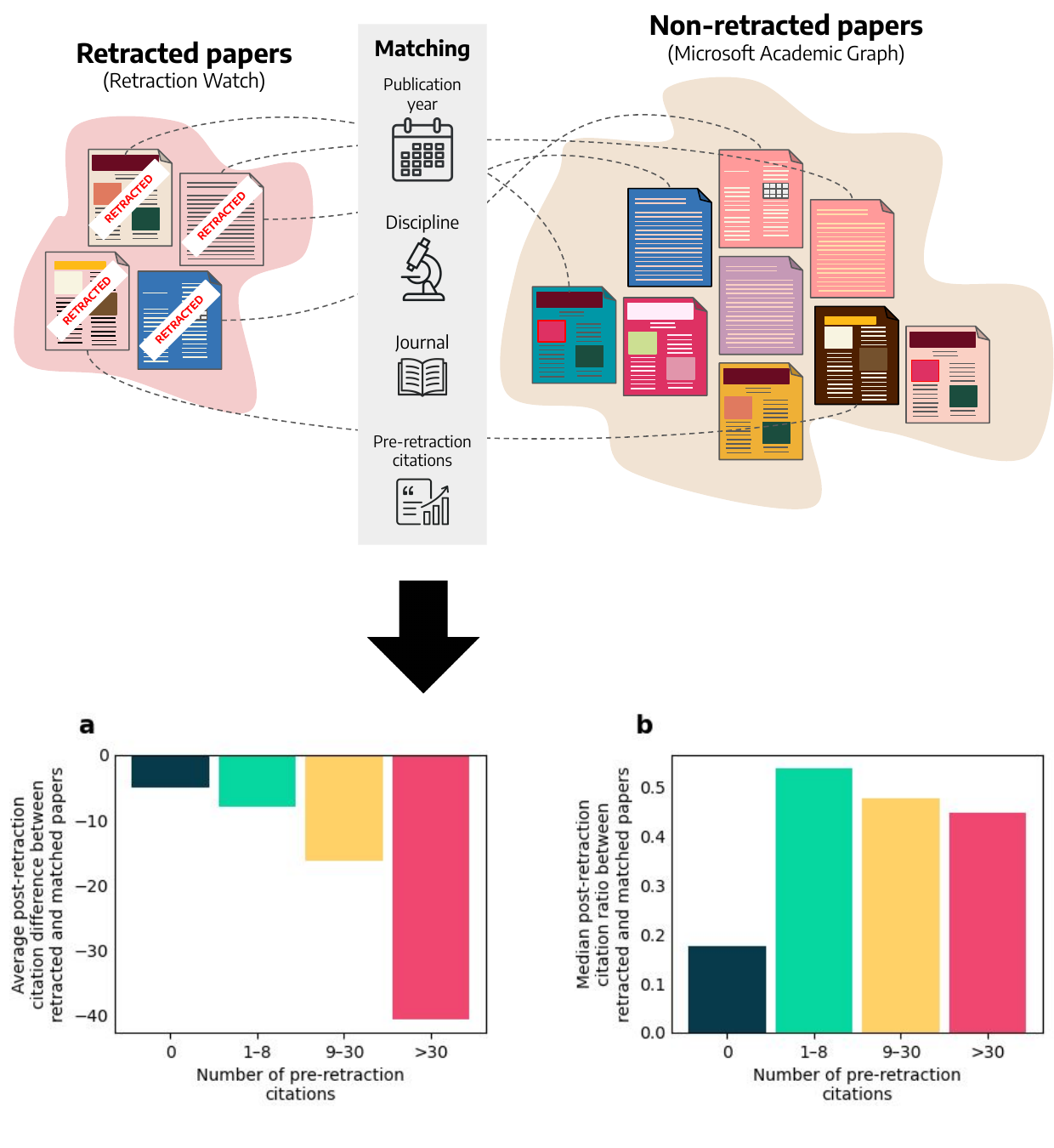}
\caption{
\textbf{Highly cited retracted papers experience greater drops in post-retraction citations.}
\textbf{(a)} Absolute post-retraction citation difference between each retracted paper and its matched control (outcome 1).
\textbf{(b)} Post-retraction citation ratio between retracted and matched papers (outcome 2).
}
\label{fig:percentdrop1}
\end{figure}

Regarding P1, for outcome 1, we compute the difference in post-retraction citations between retracted and matched papers across all citation groups. As evident in Figure \ref{fig:percentdrop1}a, there is a clear and significant trend: more highly cited papers experience larger drops in absolute citation counts. These results are significant on an aggregate level ($p<0.001$) using omnibus Welch Anova. All post-hoc pairwise comparisons between citation groups are also significant ($p<0.001$) using Welch t-tests. 

For outcome 2, we measure the ratio of the post-retraction citation counts between retracted and matched papers. We employ median as our summary statistic for this comparison. As can be seen in figure~\ref{fig:percentdrop1}b, except for the papers with 0 citations, there is a consistent drop in post-retraction citation ratio as we move from low-to-high cited papers. The difference in citation groups is found significant on an aggregate level using Kruskal–Wallis test ($p<0.001$). Pairwise Dunn's tests with Holm adjustment show that the 0-citation group differs significantly from all others ($p<0.001$). The 1–8 citation group differs from the $>$30 group ($p = 0.03$), but not from the 9–30 group ($p = 0.29$), and there is no significant difference between the 9–30 and $>$30 citation groups ($p = 0.29$). See our online appendix for more details related to the summary statistics related to the two outcomes.

\begin{figure}[!htbp]
\centering

\begin{subfigure}{\textwidth}
  \centering
  \includegraphics[width=\linewidth]{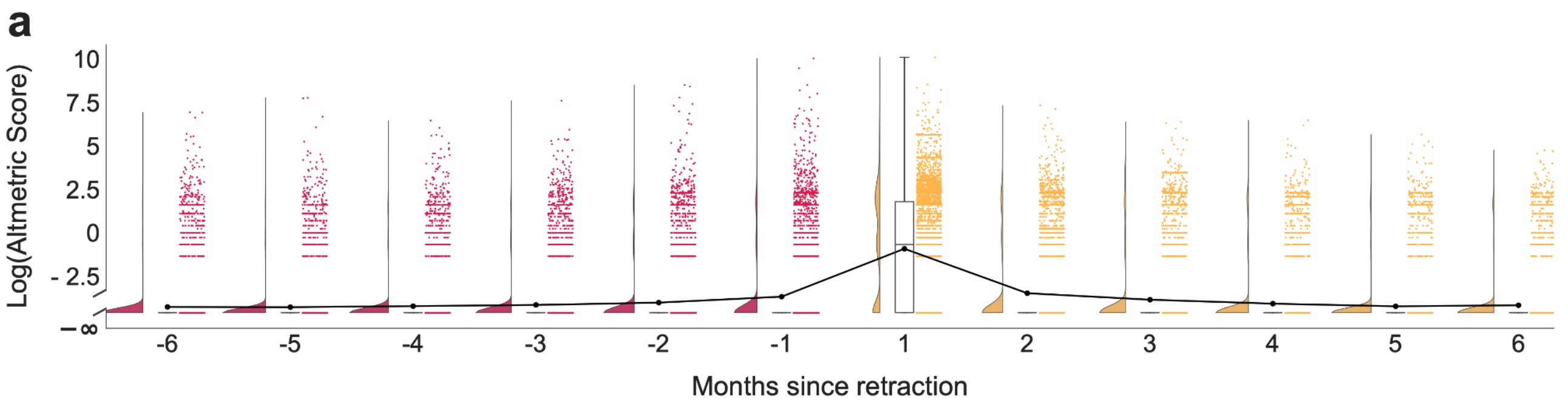}
\end{subfigure}

\vspace{1em} 

\begin{subfigure}{\textwidth}
  \centering
  \includegraphics[width=\linewidth]{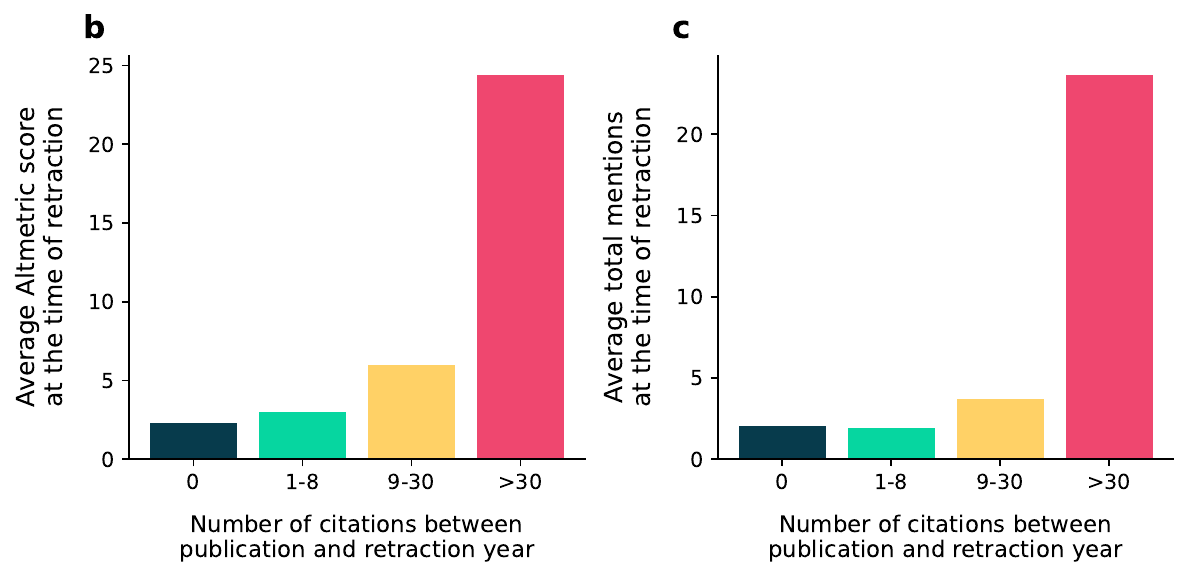}
\end{subfigure}

\caption{
\textbf{Highly cited retracted papers receive more attention during retraction.}
\textbf{(a)} shows distribution of log-transformed Altmetric scores in the 6 months before and after retraction. The x-axis represents monthly time windows relative to the retraction date, where 0 corresponds to the month of retraction (omitted for clarity), -1 indicates the month immediately preceding retraction, and +1 the month immediately following. The y-axis displays the log-transformed Altmetric score for each paper within a given month. \textbf{(b)} and \textbf{(c)} respectively show the average Altmetric score and the average number of mentions at the time of retraction for different citation groups.}
\label{fig:attention}
\end{figure}

Regarding P2, we reproduce figure~\ref{fig:attention}a from \citet{memon2025characterizing} showing that Altmetric scores peak around the first month of retraction. We isolate a 12-month `retraction window' and characterize how attention varies across different citation groups within that time window. As figure \ref{fig:attention}b and \ref{fig:attention}c show, the raw Altmetric mentions, and Altmetric scores were much higher for the more cited papers. This effect is especially notable in the most highly cited papers. We further control for various confounders, as detailed in the methods section, and estimate a linear regression model. We find that papers with more pre-retraction citations are significantly more likely to receive retraction-related attention and mentions. More concretely, each additional pre-retraction citation is associated with a 0.168 unit increase ($p<0.001$) in the raw altmetric score, and a 0.195 unit increase ($p<0.001$) in the number of mentions, all else equal. More recent publication years are positively associated with both outcomes, while other factors such as number of authors, journal rank, and time between publication and retraction show no significant effects.  Details related to regression available in our online appendix.
 
\section{Discussion}
\label{conclusion}

We take our data analysis to largely support the prediction that more widely known findings are easier to retract.  As noted, we see more significant drops in total citation rates post-retraction for highly cited papers, indicating that a larger number of authors who would have cited these papers do not.  We see the same pattern with relative drop in citation rate, though some comparisons in this analysis are not significant.\footnote{And as noted, this trend does not hold for papers with zero citations, but we assume retraction before any sort of community uptake yields different dynamics.}

In addition, more highly cited papers receive more attention during retraction.  The observation that people are more likely to share relevant information drives modeling assumptions in \citet{lacroix2021dynamics}, and also the expectation that more attention will be paid to retractions of widely known results.  In this way, the Altmetric data further validates model predictions.

One limitation is our dependence on citation count as a proxy for 1) familiarity with and 2) trust in a scientific result and its retraction.  Of course, there are cases where authors know a result, but do not cite it.  In studying retraction this may be especially relevant, since there is evidence that scientists are good at predicting which findings replicate \citep{holzmeister2025examining}.  This might mean that pre-retraction citation counts are systematically impacted by the very features that lead to retraction---that a study is fraudulent, or low quality, or contains significant errors.  In addition, sometimes researchers cite papers they know to be retracted, including in cases of self-citation \citep{madlock2015lack}.  So citation drop may not perfectly correlate with exposure to the fact of retraction.

Another limitation is our dependence on comparisons to matched papers.  This allows us to approximate what citations we might expect if the target papers had not been retracted. This said, these matches may not be perfect predictors for future citation count of retracted papers. We account for a set of covariates (e.g., publication year, journal, discipline, citations), but there may be other covariates critical to this matching, and, even if we are accounting for the key covariates, there could be imbalances that lead to biased estimates. In addition for obvious reasons we cannot match on the fact of retraction itself.  Papers that are eventually retracted presumably differ systematically from those which are not, on features like the use of fraud, poor quality methods, and error.  These differences might have led to systematic differences in future citations even if the relevant papers had not been retracted.

One last issue is that our dataset does not differentiate approving citations from ones that acknowledge retraction, or even are about retraction.  There may be systematic differences in the degree to which widely-known papers are cited as examples of retraction.

There are some relevant policy take-aways.  First, it might be important to more actively spread information about retraction of less influential work.  Previous authors have advocated that journals, pre-print servers, and scientific search engines do more to communicate about retraction.  We support these reforms, and point to a place where they may be especially important---in cases where information about the retraction itself is less likely to spread endogenously.  

Second, it may be effective to increase the relevance of retractions themselves. Packaging a retraction with information about why it may be important to an area of study may help it spread.  For instance, journals issuing a retraction, or science journalists writing about it, might explain how claims in the original paper can potentially lead future researchers astray.

A more general take-away is that in thinking about retraction policy, dynamics of social information spread should be taken into account. These dynamics seem to impact whether and how researchers receive information about retraction.

This project also has take-aways related to the combination of diverse methods in the study of philosophy of science and metascience, and related to the use of simple social models. The models presented by \citet{lacroix2021dynamics} do not provide sufficient evidence to support policy proposals, but do direct attention to an unexpected phenomenon---the possibility that high impact work may be easier to retract. This possibility has not been previously studied, and without the models would not have been a natural hypothesis. The integration of models and data analysis, then, was crucial to this investigation.  Our results find support for the models, and also clarify the relationship between prominence of a result and success in retracting it.

\section*{Acknowledgments}
We thank the PSA 2024 audience, and fellow symposiasts. Thanks to
Travis LaCroix and Anders Geil for their modeling work. We thank the Knight Foundation for funding support to S.M. and J.D.W. We thank the High Performance Computing Center at NYU Abu Dhabi for their technical support and resources. We are grateful to Ivan Oransky, Adam Marcus, and The Center for Scientific Integrity for maintaining and sharing Retraction Watch data. We also acknowledge Altmetric.com and Microsoft Academic Graph for providing data for our study. Finally, many thanks to Bedoor AlShebli and Kinga Makovi for access to their processed data as well as their feedback.

\section*{Funding Statement}
S.A.M. and J.D.W. receive funding from the John S. and James L. Knight Foundation (G-2019-58788) and the UW Center for an Informed Public that supported this research.

\section*{Declarations}
J.D.W. is on the Board of Consensus.app.  

In addition J.W. received the following restricted funds in the 36 months preceding submission of this paper:

Institute of Museum and Library Services (IMLS). Misinformation Media Literacy: Supporting Libraries as
Hubs for Misinformation Education (LG-255047-OLS-23)

NIH: Title: Developing user-centric training in rigorous research: post-selection inference, publication bias,
and critical evaluation of statistical claims. (5UE5NS132949-03)

NSF: Convergence Accelerator Track F: Co-Designing for Trust: Reimagining Online Information Literacies
with Underserved Communities. Phase II (2230616)

NSF: Collaborative Research: SaTC: CORE: Large: Rapid-Response Frameworks for Mitigating Online
Disinformation (2120496)

\bibliographystyle{psalike-v1.0}   
\bibliography{retraction}
\end{document}